\documentclass{article}
\usepackage{spconf,amsmath,graphicx}
\usepackage{booktabs,tabularx, subcaption, float}
\title{Late reverberation suppression using U-nets}
\name{
Diego León$^{\star }$ 
\qquad 
Felipe Tobar$^{\dagger, \ddagger}$
}
  
\address{
$^{\star}$Department of Electrical Engineering, Universidad de Chile \\
$^{\dagger}$Initiative for Data \& Artificial Intelligence, Universidad de Chile \\
$^{\ddagger}$Center for Mathematical Modeling, Universidad de Chile
}
\begin{document}
\ninept
\maketitle
\begin{abstract}
In real-world settings, speech signals are almost always affected by reverberation produced by the working environment; these corrupted signals need to be \emph{dereverberated} prior to performing, e.g.,  speech recognition, speech-to-text conversion, compression, or general audio enhancement. In this paper, we propose a supervised dereverberation technique using \emph{U-nets with skip connections}, which are fully-convolutional encoder-decoder networks with layers arranged in the form of an ``U'' and connections that ``skip'' some layers. Building on this architecture, we address speech dereverberation through the lens of Late Reverberation Suppression (LS). Via experiments on synthetic and real-world data with different noise levels and reverberation settings, we show that our proposed method termed ``LS U-net'' improves quality, intelligibility and other  performance metrics compared to the original U-net method and it is on par with the state-of-the-art GAN-based approaches.

%We present two architectures building on this concept: the first %one considers generative adversarial networks and uses the U-net %as generator. In the second one, inspired by applications %involving Long Short Term Memory (LSTM) networks, we use the %U-net for Late Reverberation Suppression. Our experimental %results, carried out on different noise levels and reverberation %settings, show that both proposed architectures improve quality, %intelligibility and other dereverberation performance metrics %compared to the original U-net method.\\  

\end{abstract}
%%%%%%%%%%%%%%%%%%%%%%%%%%%%%%%%%%%%%%%%%%%%%%%%%%%%%%%%%%%%%%%%%%%%
%%%%%%%%%%%%%%%%%%%%%%%%%%%%%%%%%%%%%%%%%%%%%%%%%%%%%%%%%%%%%%%%%%%%
\section{Introduction}
\label{introduction}
Speech reverberation is an acoustic phenomenon whereby reflections of the acoustic signal (over surfaces and objects) are combined with the original signal at the receiver's end. The resulting \emph{reverberated} signal is thus a corrupted one, where the  intelligibility and quality of the speech is degraded \cite{vejestorio}. Perceived reverberation levels depend on a number of factors, including the geometry of the room, the materials used in it, and the distance between the speaker and the receiver \cite{naylor1}.

Reverberation can be modelled as a convolution between a source signal and the room impulse response. Based on this modelling assumption one can design dereverberation techniques to recover the source (original) signal from observations of the received (convolved) signal. A popular unsupervised approach is the Weighted Prediction Error (WPE) \cite{wpe} method. WPE estimates the original signal by applying a linear filter to the received signal, where the filter, learnt via maximum likelihood, assumes a Gaussian prior on the source signal (possibly heteroscedastic  \cite{wpe_neural}). There are several extensions of WPE, in particular, the frequency domain normalized delayed linear prediction (FD-NDLP) method \cite{fdndlp} is an efficient implementation of WPE which uses the short-time Fourier transform (STFT) and is known to outperform its temporal-domain counterpart.

Deep learning has also been recently used in speech dereverberation. For instance, multilayer perceptrons (MLP) and long short-term memory (LSTM) networks have been developed to learn mappings from a window of reverberated frames (or ``context'' windows) to a source frame, thus \emph{learning to dereverberate} by inverse transformations \cite{dnn,dnn2,becerra}. Additionally, Zhao et al.~\cite{late_dev} proposed an LSTM-based late-reverberation-suppression strategy which learns the difference between the source and reverberated signals, therefore, dereverberation is performed by substracting the late reverberation estimation to the observed reverberated signal. 

Architectures using deep autoencoders have too been considered for audio generation \cite{synthesis} and in particular for dereverberation \cite{derev_autoencoder}, while generative adversarial networks (GAN) have been shown to improve training for some dereverberation methods \cite{audio_gan1,gan_derev}. Building on these tools, Ori Ernst et al.~\cite{unet} used an encoder-decoder fully convolutional neural network called U-net (due to its layers arranged in the shape of an ``U'' \cite{unet_orig}) for speech dereverberation. Their strategy was to learn the mapping between the (log) power spectrum between the reverberated and source signals as if they were images. In the same work, Ori Ernst et al. used a U-net as generator in a GAN.

In this work, we propose a novel U-net architecture for speech dereverberation. The unique feature of the proposed method is that it implements the U-net in a \emph{Late Reverberation Suppression} (LS) setting, while in previous works i) LS has been addressed using LSTMs \cite{late_dev}, and ii) U-nets have been used for direct reverberation \cite{unet} (and not for LS). Our proposed method exhibits significantly better results than traditional U-net in terms of popular intelligibility, quality and reverberation objective measures (e.g., speech-to-reverberation modulation energy ratio, SRMR), and achieves dereverberation indicators that are similar to recent extensions of the U-net architecture trained using GANs.
%%%%%%%%%%%%%%%%%%%%%%%%%%%%%%%%%%%%%%%%%%%%%%%%%%%%%%%%%%%%%%%%%%%%
\section{Problem Formulation}
\label{problem}

Let $x(\cdot)$ be the source signal and $y(\cdot)$ the reverberated signal given by the convolution between the source and a room impulse response (RIR) $h(\cdot)$. Let us also consider the reverberation time $T_{60}$, given by the time it takes for a signal to decay 60 dB relative to the level of direct sound (initial impulse) \cite{naylor1}. The reverberation time $T_{60}$, uniquely determined by $h$, is relevant since it is a measure ``how reverberant'' a signal is when is convolved with $h(t)$. For instance, a reverberation time $T_{60}=0.2s$ represents a low level of reverberation, while $T_{60}=0.6s$ produces a noticeable reverberation level. 

By considering a source of additive noise $\eta(\cdot)$, the model relating the above defined objects is given by 
\begin{equation}
\label{eq:conv1}
    y(t) = (x*h)(t) + \eta(t),
\end{equation}
where ``$*$'' denotes convolution operator. Dereverberation is thus defined as a \emph{blind deconvolution}, that is, the task of recovering $x(t)$ using observations of $y(t)$ in eq.~\eqref{eq:conv1} when the $h(t)$ is unknown. Notice that by splitting the room impulse response $h(t)$ in \emph{early reflections} $h_{\text{early}}(t)$ and \emph{late reflections} $h_{\text{late}}(t)$, eq.~\eqref{eq:conv1} can be expressed as 
\begin{equation}
\label{eq:rev}
   \begin{split}
    y(t) &= (x*h_{\text{early}})(t) + (x*h_{\text{late}})(t) + \eta(t)\\
    &= y_{\text{early}}(t) + y_{\text{late}}(t),
   \end{split}
\end{equation}
where $y_{\text{early}}$ is as close as possible to the desired (source) signal $x$, since the reverberation is largely due to late reflections. 

The problem on which this work will focus is that of suppressing the late reflections $y_{\text{late}}(t)$. 

%Yan Zhao et al. \cite{late_dev} showed through objective quality %and intelligibility measures, for several reverberation times, %that late reverberation suppression is a powerful dereverberation %method using Long Short Term Memory (LSTM). \textbf{cerrar esta %seccion de otra forma: o bien no decir nada de Zhao, o bien %decirlo y decir qué vamos a hacer nosotros de forma distinta}

%%%%%%%%%%%%%%%%%%%%%%%%%%%%%%%%%%%%%%%%%%%%%%%%%%%%%%%%%%%%%%%%%%%%
\section{Proposed method and baselines}
\label{baseline}

\begin{figure}[t]
\centering
\includegraphics[width=0.5\textwidth]{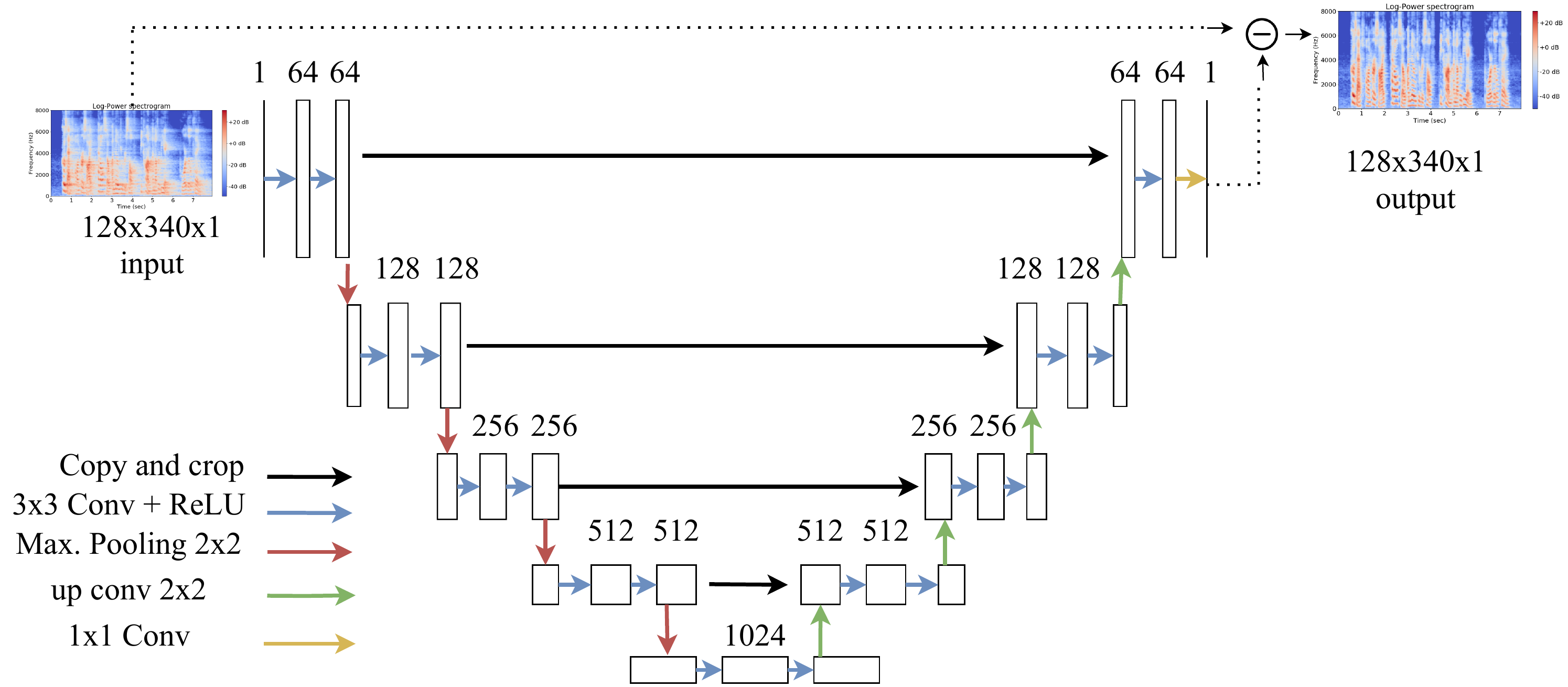}
\caption{Proposed U-net architecture for speech dereverberation. Skip connections are represented by horizontal continuous black arrows and the dashed connection is the distinguishing feature of our method that enables it use for late reverberation suppression.}
\label{supunet}
\end{figure}

%tengo la duda si dejar o no este parrafo, concuerdo que es medio redundante

%A U-net is a fully convolutional neural network with %encoder-decoder structure originally developed for image %segmentation \cite{unet_orig}, its name is inspired on the ``U'' %shape followed by the layers of the net, which are %``skip-connected'', meaning that encoder convolutional blocks are %directly connected to decoder convolutional blocks thus skipping %central layers. \cite{unet} proposed a U-net for dereverberation %whereby the log magnitude spectrogram was fed to the net as if it %were an image.

Our proposal extends previous works in the literature by focusing on the problem of late reverberation suppression (LS) considered in \cite{late_dev} (originally addressed using LSTMs) with the U-net architecture presented in \cite{unet_orig}; originally for plain dereverberation. Figure \ref{supunet} shows the U-net architecture proposed in our work. 

The main difference between our contribution and previous methods is the skip connection between input and output (dashed line in Figure \ref{supunet}), which is not present in the original dereverberation U-net \cite{unet_orig}. This skipped connection is what materialises our focus on late reverberation suppression (rather than plain dereverberation): the proposed U-net architecture does not learn a mapping from reverberated to dereverberated spectrogram, but instead it learns to generate an intermediate log-power-spectrum ``image'', which is subtracted of the input (observed reverberated signal). 

The intuition supporting the proposed architecture follows the idea that estimating the late reflections $y_{\text{late}}(t)$ is simpler than estimating the full reverberated signal $y(t)$, since it is known that the true signal does appear as a component in the reverberated signal---see eq.\eqref{eq:rev}. Therefore, by giving the U-net a less challenging task (by placing the input-output skipped connection shown at the top of Figure \ref{supunet}) our hypothesis is that the proposed U-net architecture will have improved performance in reconstructing the source signal against over the baseline U-net in \cite{unet_orig}. This is because the baseline U-net aims to learn the  reverberant-dereverberant mapping without any prior knowledge of the dereverberation process, in particular it does not considers that the source appears as an early reflection. The proposed model is trained in the same fashion as the baseline U-net using the MSE loss function. 

For purposes of experimental validation, we will consider a recently proposed dereverberation method \cite{unet} based on a Generative Adversarial Network (GAN) using a U-net as generator, this architecture is known to improve the quality of dereverberated spectrograms generated over the original U-net method in \cite{unet_orig}. In this method, the discriminator network classifies between the generator output spectrogram and the clean spectrogram or ``target''. Learning using this strategy uses the following loss function:
\begin{equation}
    \label{eq:gan1}
    L(G, D) = L_{\text{GAN}}(G, D)+ \lambda L_{\text{MSE}}(G),
\end{equation}
where $L_{\text{MSE}}(G)$ is the mean square error between the generator output and the target log power spectrum, $\lambda$ is an hyper-parameter controlling the MSE weight in the loss function and $L_{\text{GAN}}(G, D)$ has the traditional form of GAN loss.

In addition to U-net-based architectures above, we consider other known methodologies to dereverberation, mainly based on MLP and LSTMs. Summarising, all the architectures to be implemented in our experiments are

\begin{itemize}
    \item \textbf{LS U-net}: Proposed Late Reverberation Suppression U-net
    \item \textbf{U-net}: The original U-net method, a symmetric U-net structure for dereverberation \cite{unet_orig} trained on an MSE loss
    \item \textbf{U-net GAN}: a GAN architecture using a symmetric U-net as generator \cite{unet}
    \item \textbf{Context-MLP}: An MLP with Context Window \cite{dnn}\cite{dnn2}
    \item \textbf{Context-LSTM}: An LSTM with Context Window \cite{becerra}
    \item \textbf{LS-LSTM}: A late reverberation suppression LSTM \cite{late_dev} 
    \item \textbf{FD-NDLP}: The frequency domain  normalized  delayed  linear  prediction  \cite{fdndlp} (which is unsupervised)
\end{itemize}

All the above architectures were implemented exclusively for our experiments with the exception of FD-NDLP, for which we relied on officially released code. Training was performed using Adam \cite{kingma2015adam} and a batch size of 16. U-net GAN, in particular,  was trained using $\lambda=\text{1e-2}$, chosen experimentally in order to keep MSE and $L_{\text{GAN}}$ in the same magnitude order. Furthermore, the input log power spectrogram for U-net GAN \textbf{was not normalized}, but we set a minimum value of -80 dB and a maximum of 30 dB; this was because our preliminary results exhibited poor performance using  normalization to confine the input in the [-1, 1] range or confining the output in the same range using tanh($\cdot$). 

%The non-U-net method above work using frames or a small number of them as input, then they can be compared with a different approach using convolutional networks (U-net in the 3 variants).
%We used MLP and LSTM with Context Window \cite{dnn}\cite{dnn2}\cite{becerra},  Late Reverberation Suppression LSTM (We abbreviate it as LS-LSTM)\cite{late_dev}, and the unsupervised method FD-NDLP \cite{fdndlp}. 

%We implemented U-net in 3 variants: U-net (baseline) \cite{unet}, U-net GAN \cite{unet} and the proposed Late Reverberation Suppression U-net. We denote the last model as LS U-net (LS for Late Suppression). 

%Esta figura mostraba un esquema de GAN, pero creo que sobra si es un metodo ya existente y no propuesto

%\begin{figure}[t]
%\centering
%\includegraphics[width=3.0in]{Images/gan_esquema.pdf}
%\caption{Generative Adversarial Network scheme for speech %dereverberation. Generator is U-net architecture in this case.}
%\label{gan_scheme}
%\end{figure}

%%%%%%%%%%%%%%%%%%%%%%%%%%%%%%%%%%%%%%%%%%%%%%%%%%%%%%%%%%%%%%%%%%%%

\section{Experiments}
\label{experiments}

\subsection{Datasets and pre-processing}

Our experiments considered synthetic and real-world data. The former were taken from the LibriSpeech \cite{librispeech} database, whose utterances (audio examples in dataset) are sampled at 16 kHz. Our procedure to generate the  synthetic reverberated speech was by convolving the LibriSpeech audio signals with RIR from the Omni \cite{omni} and MARDY \cite{mardy} databases. The real-world data considered in our experiments came from the BUT Speech@FIT Reverb Database \cite{real_data}, which are retransmitted signals also taken from LibriSpeech, and are thus naturally reverberated.

Spectrograms, in all cases, were computed using FFT with a  window length of 2048 samples and \emph{hop} length of 512 samples; a Mel filterbank was used to reduce the bin size. Experimentally, we chose between 128 and 256 bins, in both cases it was possible to recover the temporal signal appropriately but when using a smaller number of bins (e.g., 64 bins) the signal was recovered with difficulty. Lastly, we used 128 bins and set the number of frames for each spectrogram to 340 (which was the mean value of frames over all training spectrograms) using Lanczos interpolation available on OpenCV.

\subsubsection{Simulated data}

RIRs from databases  
Omni \cite{omni} and MARDY \cite{mardy} were used to generate reverberant speech audios. Omni is composed of 3 rooms, 2 of which were used for training and the remaining one for testing. MARDY (1 room) was used for testing only.

The \textbf{reverberant training data} was produced using random RIR utterances and random SNRs chosen from the range [15, 35] dB for each example. This strategy allowed for a training set with a wide variety of noise and reverberation time. Reverberation time varied in an approximate range of 0.3s and 0.7s for the considered databases. The \textbf{reverberant test data} was generated using Omni RIRs dataset for $\text{SNR}= 15\text{dB}$ and $\text{SNR}= 35\text{dB}$ (the same 500 utterances for each SNR). Another 500 utterances were produced using the MARDY RIRs dataset for near and far microphones, where noise was fixed at $\text{SNR}= 35\text{dB}$. 

These synthetic signals were produced using the RIR generator\footnote{Available on https://pypi.org/project/rir-generator/}, based on the original method proposed by \cite{allen_79}, in order to introduce $T_{60}$ variability. Ths way, the simulated data were generated for $T_{60}$ varying between 0.2 and 1.0s (9 values spaced in 0.1s) and using 50 utterances for each $T_{60}$ value.

\subsubsection{Re-transmitted real-world data}

We used the BUT Speech@FIT Reverb Database \cite{real_data}, which contains LibriSpeech re-transmitted for near and far microphones. We used 500 utterances for near and far microphones. Our quantitative evaluation was based on the following metrics:\\

\noindent $\bullet$ \textbf{PESQ}: Perceptual Evaluation of Speech Quality \cite{pesq}\\
\noindent \quad $\bullet$ \textbf{CD}: Cepstral Distance\\
\noindent \quad $\bullet$ \textbf{LLR}: Log-Likelihood Ratio\\
\noindent \quad $\bullet$ \textbf{fwSNRseg}: Frequency Weighted Segmental SNR  \cite{metrics1}\cite{metrics4}\cite{book_metrics}\\
\noindent \quad $\bullet$ \textbf{SRMR}: Speech to Reverberation Modulation Energy Ratio \cite{metrics2}\\

The first four metrics are \emph{intrusive metrics}, which compare the input signal with a clean signal (in terms of reverberation and noise) and then provide ``similarity'' scores. The SRMR metric, on the contrary, is a representation obtained by means and auditory-inspired filterbank (based on the functioning of the cochlea) analysis of critical band temporal envelopes of the speech signal \cite{metrics2}. Using this last non-intrusive measure is relevant regarding the realistic evaluation of the methods considered, since in real-word applications clean signals that can be used as benchmarks may not available.

\subsection{Results for synthetic data: varying noise}

\begin{table*}[t]
\footnotesize
   \centering
   \setlength\tabcolsep{8pt} % default: 6pt
   \caption{Results of simulated data for $\text{SNR}=15\textbf{dB}$ y $\text{SNR}=35\textbf{dB}$  using Omni RIRs dataset. ($\uparrow$): higher is better, ($\downarrow$): lower is better.} 
   \begin{tabular}[t]{@{} l *{10}{c} @{}}

   \cmidrule(l){1-11}
   & \multicolumn{2}{c}{PESQ ($\uparrow$)} 
   & \multicolumn{2}{c}{CD ($\downarrow$)}  
   & \multicolumn{2}{c}{LLR ($\downarrow$)}
   & \multicolumn{2}{c}{fwSNRseg ($\uparrow$)}
   & \multicolumn{2}{c@{}}{SRMR ($\uparrow$)} \\
   \cmidrule(lr){2-3} \cmidrule(lr){4-5} \cmidrule(l){6-7} \cmidrule(l){8-9} \cmidrule(l){10-11}
   \text{SNR (dB)} $\longrightarrow$ & 15 & 35 & 15 & 35 & 15 & 35 & 15 & 35 & 15 & 35 \\
   \midrule
   Reverberant & 1.98 & 2.11 & 7.30 & 5.39 & 1.36 & 0.81 & 6.38 & 7.69 & 3.08 & 3.17 \\
   Context-MLP & 1.66 & 2.18 & 6.84 & 4.16 & 1.38 & 0.59 & 5.04 & 7.74 & 2.06 & 3.94\\
   Context-LSTM  & 1.68 & 2.31 & 6.73 & 3.91 & 1.36 & 0.52 & 5.20 & 8.42 & 1.93 & 4.06\\
   LS-LSTM  & 1.86 & 2.25 & 6.17 & 4.04 & 1.18 & 0.52 & 5.98 & 8.34 & 2.71 & 4.75\\
   FD-NDLP  & 2.09 & 2.43 & 7.45 & 4.30 & 1.39 & 0.54 & 6.96 & 9.66 & 4.25 & 4.47 \\
   U-net & 2.59 & 2.66 & 4.44 & 3.26 & 0.61 & 0.36 & 9.35 & 10.00 & 5.93 & 5.61\\
   LS U-net & \textbf{2.65} & \textbf{2.72} & 4.38 & \textbf{3.23} & \textbf{0.59} & \textbf{0.34} & \textbf{9.56} & \textbf{10.20} & 6.30 & 5.98\\ 
   U-net GAN & 2.62 & 2.69 & \textbf{4.37} & 3.34 & 0.60 & 0.36 & 9.15 & 9.82 & \textbf{7.18} & \textbf{6.73}\\
   \bottomrule
   \end{tabular}
   \label{simulated1}
\end{table*}

%\begin{figure}[t]
%        \begin{subfigure}[t]{.25\textwidth}
%            \includegraphics[width=1\linewidth]{Images/noise_interv%al.pdf}
%            \caption{SRMR versus SNR}
%            \label{noise_interval}
%        \end{subfigure}\hfill%   
%        \begin{subfigure}[t]{.24\textwidth}
%            \includegraphics[width=1\linewidth]{Images/t60_analisis%.pdf}
%            \caption{SRMR versus reverberation time}
%            \label{t60_interval}
%        \end{subfigure}\hfill%
%        \label{noiseAndT60}
%        \caption{Sensibility of SRMR to noise and reverberation time.}
%    \end{figure}

\begin{figure}[t]
\centering
\includegraphics[width=0.33\textwidth]{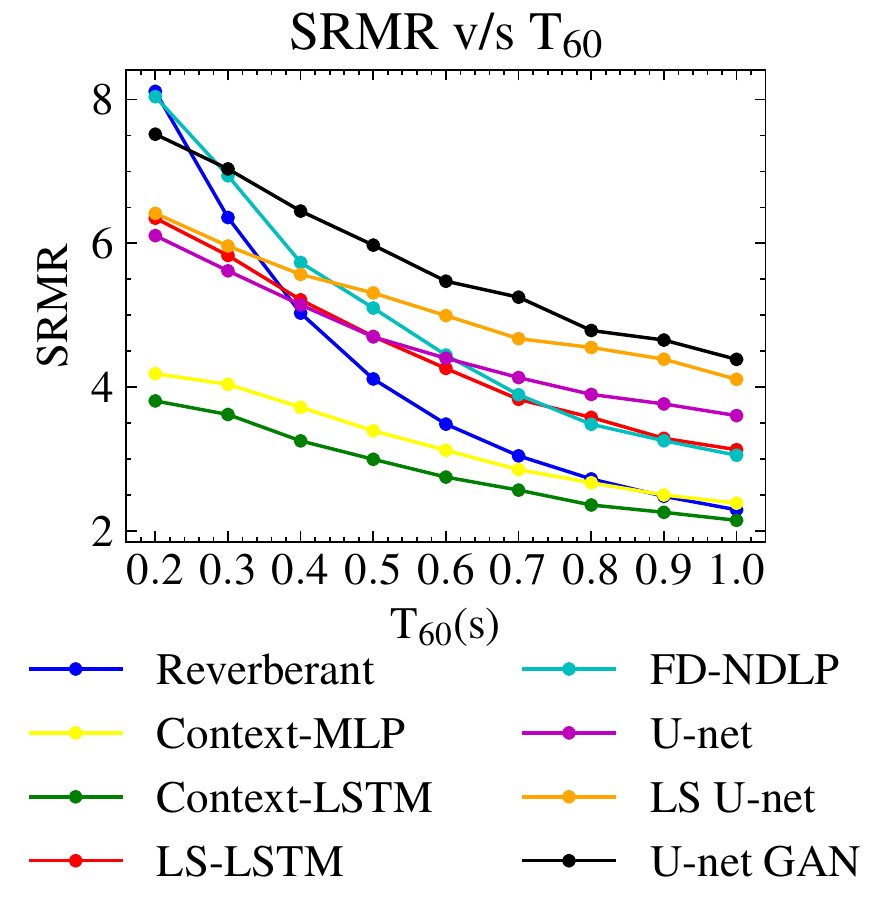}
\caption{Synthetic data: SRMR vs reverberation time.}
\label{t60_interval}
\end{figure}

Table \ref{simulated1} shows the results of simulated data for $\text{SNR}= 15\text{dB}$ and $\text{SNR}= 35\text{dB}$. The three variants of the U-net architecture exhibits the best dereverberation performance for all metrics and for both noise levels. Performances are consistent across SNR values, which shows advantages (in terms of noise) of the approaches based on U-net. The proposed LS U-net exhibits the best performance under most metrics while U-net GAN shows excels under the SRMR score, however, LS U-net still shows a clear advantage over the all other benchmarks, including the baseline U-net, for the SRMR score.

%To gain further insight into the sensibility of the proposed method %to varying noise levels, Figure \ref{noise_interval} shows mean %SRMR for SNR varying in [15,35] dB (20 values) using 50 utterances %per SNR value. U-net, LS U-net and U-net GAN show almost invariant %SRMR score in all SNR interval, which represents a robust %dereverberation behavior in terms of noise.

\begin{figure}[h!]
        \begin{subfigure}[t]{.23\textwidth}
            \includegraphics[width=1\linewidth]{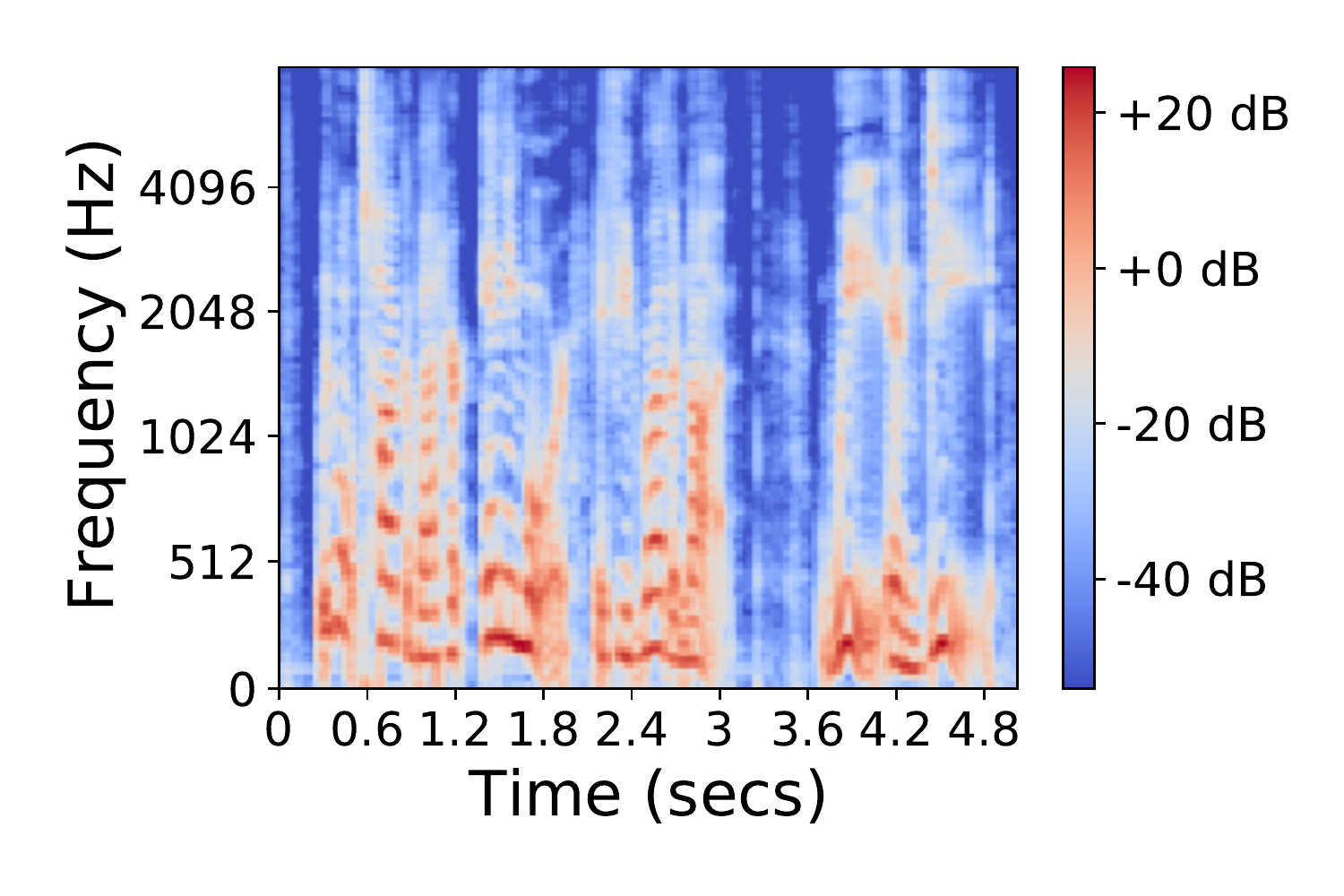}
            \caption{Clean. SRMR = 8.45.}
            \label{orig_example}
        \end{subfigure}\hfill%   
        \begin{subfigure}[t]{.23\textwidth}
            \includegraphics[width=1\linewidth]{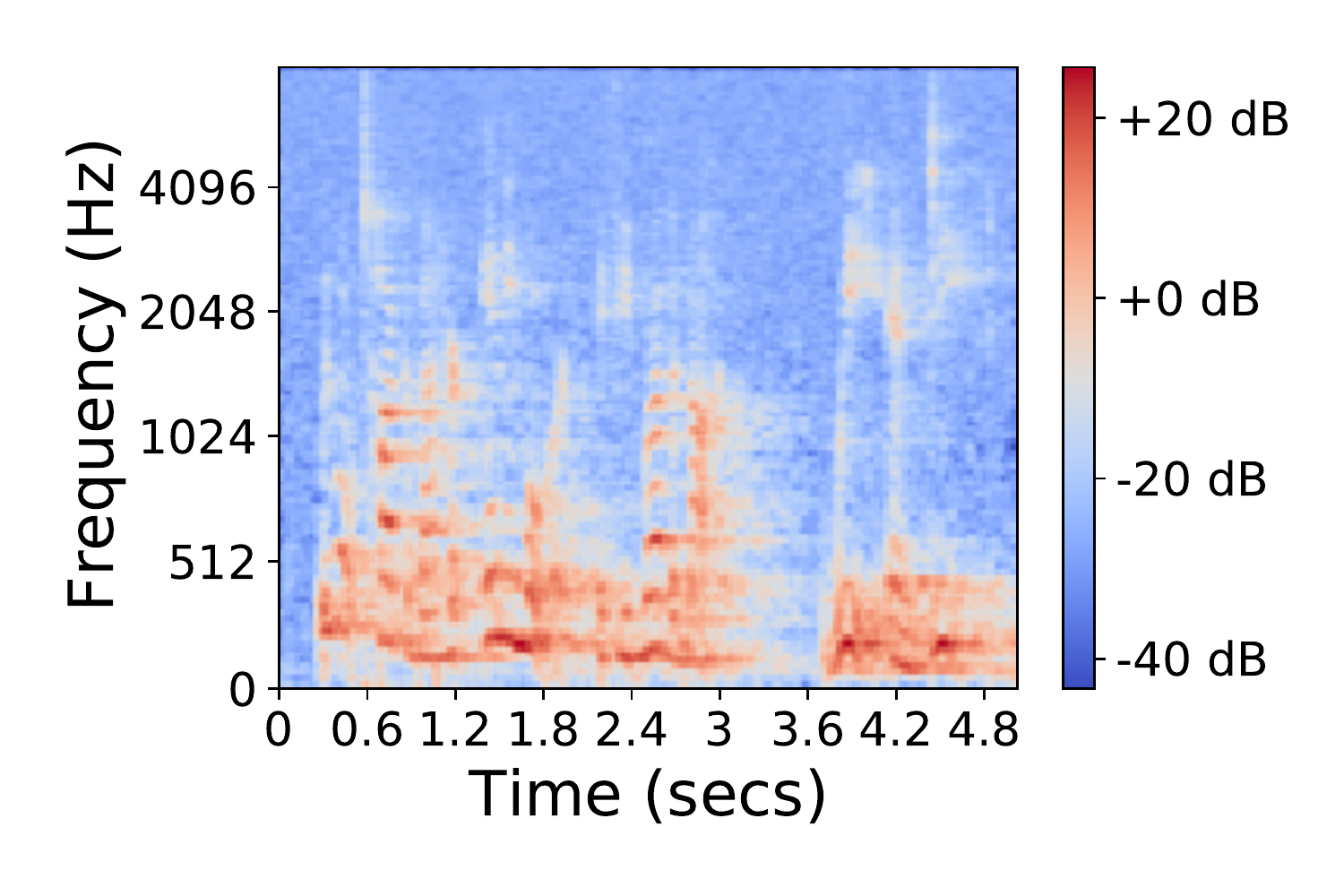}
            \caption{Reverberant. SRMR = 3.45.}
            \label{rev_example}
        \end{subfigure}\hfill%
        \begin{subfigure}[t]{.23\textwidth}
            \includegraphics[width=1\linewidth]{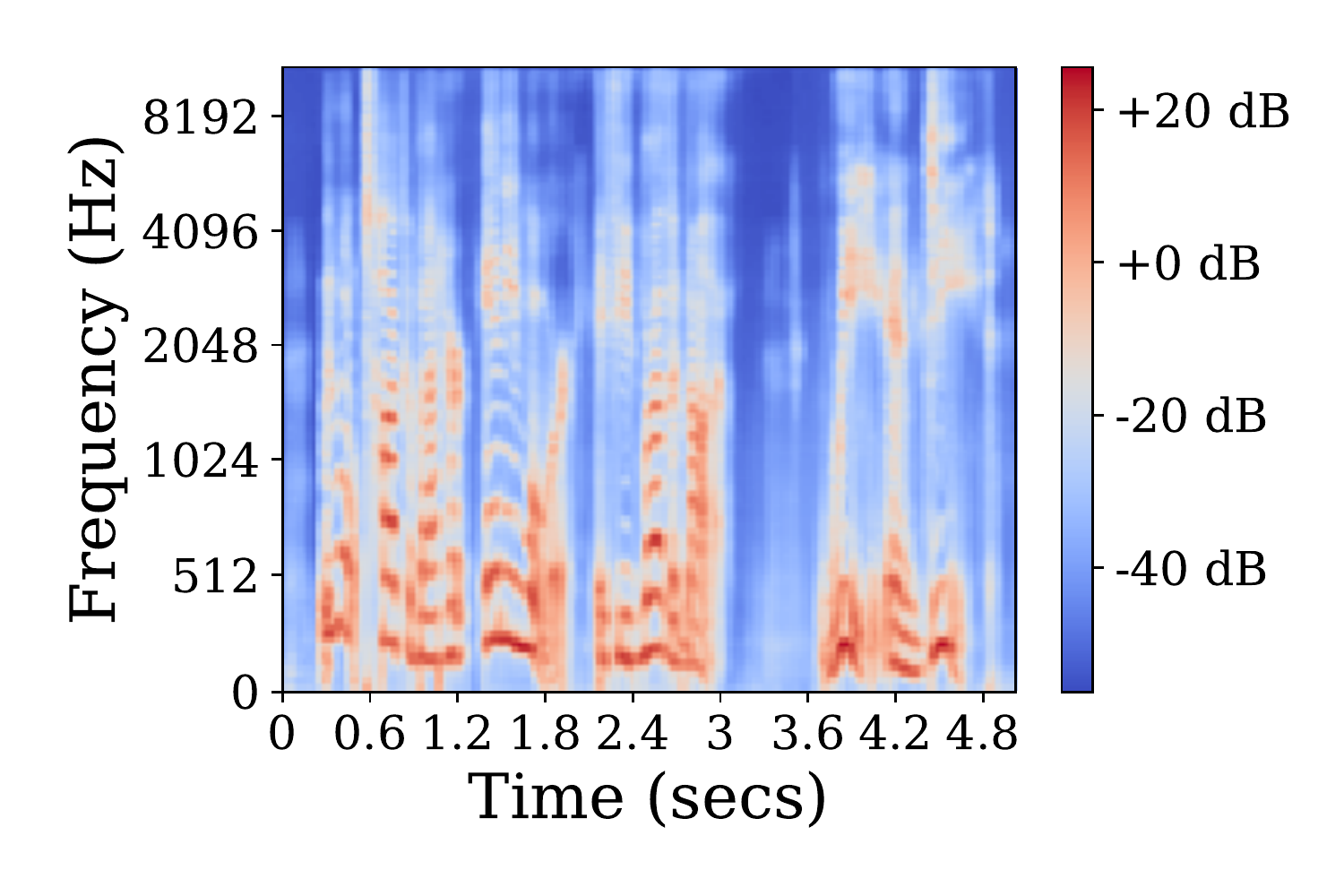}
            \caption{U-net. SRMR = 6.49.}
            \label{unet_example}
        \end{subfigure}\hfill%
        \begin{subfigure}[t]{.23\textwidth}
            \includegraphics[width=1\linewidth]{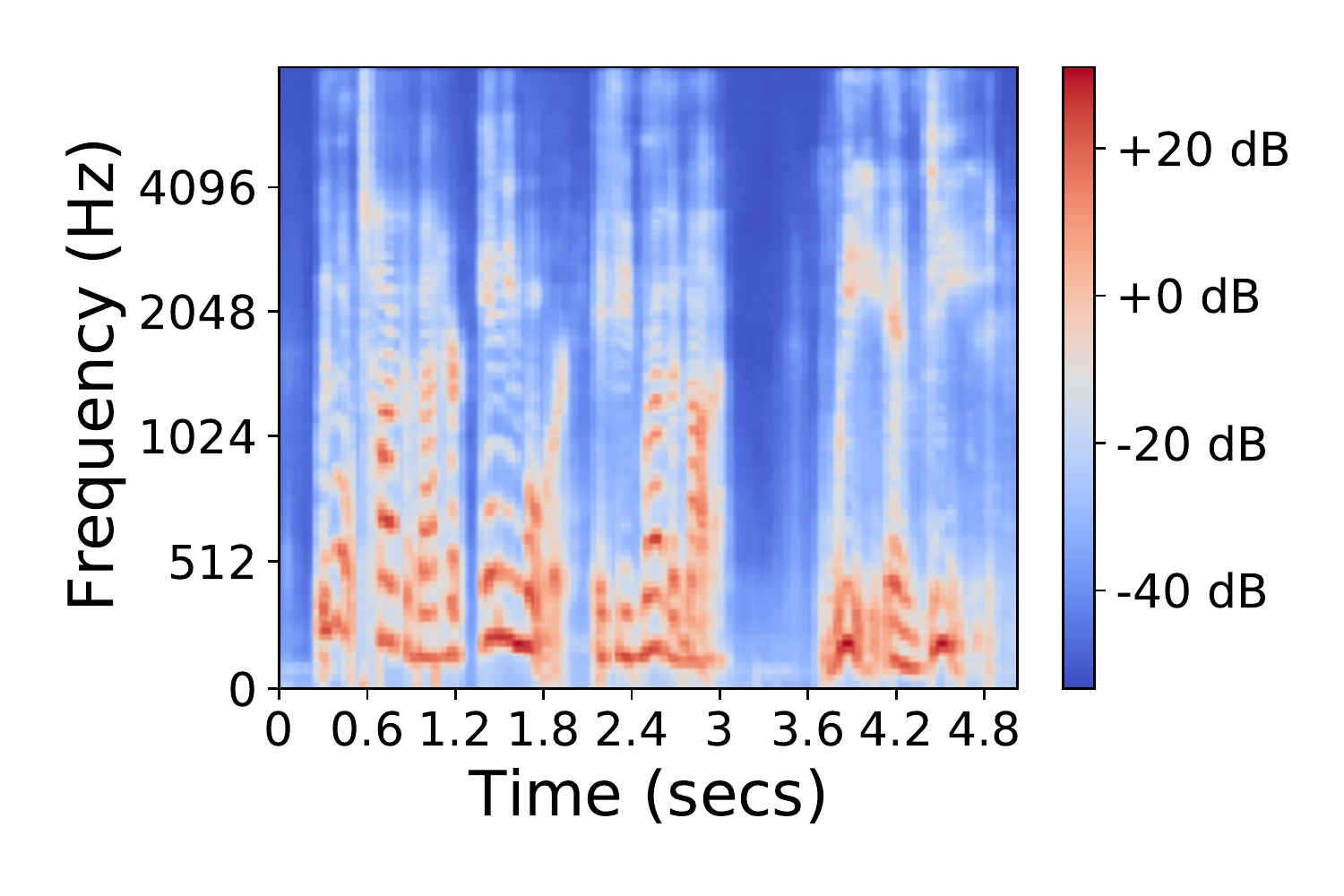}
            \caption{U-net GAN. SRMR = 7.99.}
            \label{gan_example}
        \end{subfigure}\hfill%
        \begin{subfigure}[t]{.23\textwidth}
            \includegraphics[width=1\linewidth]{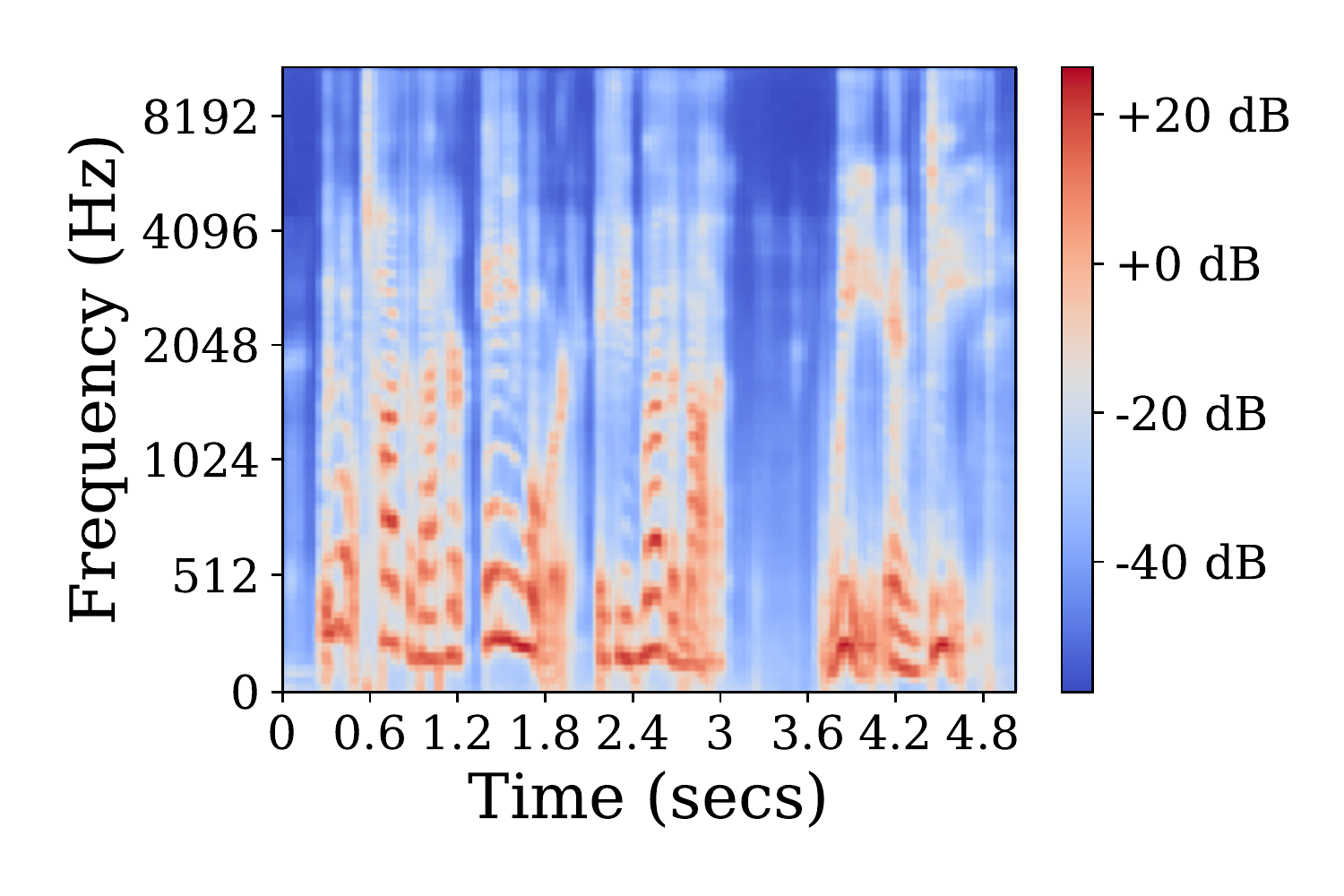}
            \caption{LS U-net. SRMR = 7.51.}
            \label{LSunet_example}
        \end{subfigure}\hfill%
        
        \caption{Example of the log power spectra for the  synthetic-data experiment.}
        \label{qual_eval}
    \end{figure}
\subsection{Results for synthetic data: varying $T_{60}$}

\begin{table*}[t]
   \centering
   \footnotesize
   \setlength\tabcolsep{8pt} % default: 6pt
   \caption{Results of simulated data using MARDY RIRs dataset. Reverberation times are 291 and 447 ms for near and far microphones respectively. ($\uparrow$): higher is better, ($\downarrow$): lower is better.}   
   \begin{tabular}[t]{@{} l *{10}{c} @{}}

   \cmidrule(l){1-11}
   & \multicolumn{2}{c}{PESQ ($\uparrow$)} 
   & \multicolumn{2}{c}{CD ($\downarrow$)}  
   & \multicolumn{2}{c}{LLR ($\downarrow$)}
   & \multicolumn{2}{c}{fwSNRseg ($\uparrow$)}
   & \multicolumn{2}{c@{}}{SRMR ($\uparrow$)} \\
   \cmidrule(lr){2-3} \cmidrule(lr){4-5} \cmidrule(l){6-7} \cmidrule(l){8-9} \cmidrule(l){10-11}
   \text{Mic. distance} $\longrightarrow$ & Near & Far & Near & Far & Near & Far & Near & Far & Near & Far \\
   \midrule
   Reverberant & 2.57 & 2.15 & 5.25 & 5.71 & 0.85 & 0.97 & 8.68 & 6.58 & 5.21 & 4.49 \\
   Context-MLP & 2.09 & 1.87 & 5.48 & 5.62 & 1.02 & 1.07 & 6.72 & 5.82 & 3.13 & 2.81\\
   Context-LSTM  & 2.14 & 1.90 & 5.49 & 5.64 & 0.99 & 1.05 & 7.21 & 6.12 & 3.05 & 2.60\\
   LS-LSTM  & 2.38 & 2.07 & 4.97 & 5.29 & 0.81 & 0.92 & 8.06 & 6.64 & 4.53 & 4.12\\
   FD-NDLP  & 2.71 & 2.24 & 5.44 & 5.83 & 0.90 & 1.00 & 8.57 & 6.60 & 6.03 & 5.34 \\
   U-net & 2.65 & 2.28 & 4.12 & 4.57 & 0.54 & 0.65 & 9.23 & 7.52 & 5.47 & 4.88\\
   LS U-net & \textbf{2.74} & \textbf{2.36} & \textbf{4.09} & 4.59 & \textbf{0.52} & \textbf{0.64} & \textbf{9.32} & \textbf{7.63} & 5.73 & 5.36\\ 
   U-net GAN & 2.72 & \textbf{2.36} & 4.11 & \textbf{4.56} & 0.53 & \textbf{0.64} & 9.24 & \textbf{7.63} & \textbf{6.60} & \textbf{6.19}\\
   \bottomrule
   \end{tabular}
   \label{simulated2}
\end{table*}
Table \ref{simulated2} shows the results of simulated data for near and far microphones. Recall that the  reverberation time $T_{60}$ (defined in Section \ref{problem}) associated to far microphones is greater than that of near microphones, this is because the reverberation effect is more subtle when the speaker is closer to the microphone. The baseline U-net certainly improved, in terms of SRMR score, for near and far microphones compared to reverberant speech and all non-U-net-based methods; however, observe that LS U-net and U-net GAN had significantly better scores overall. The unsupervised method FD-NDLP (based on Late Reverberation Suppression) was competitive for near and far microphones. The indicators PESQ, CD, LLR and fwSNRseg showed small differences among the 3 U-net variants, although LS U-net shows the best performance in most cases.

Figure \ref{t60_interval} shows SRMR results of simulated data using RIR generator as a function of $T_{60}$. The RIR generator was used assuming a room of dimensions 5[mt]$\times$4[mt]$\times$6[mt] (width, length and depth). As expected, the SRMR score decreases for increasing $T_{60}$ for all methods, with the reverberant (unprocessed, shown in blue) signal having the sharpest decay and the proposed LS U-net (orange) closely following the sate-of-the-art U-net GAN (black). None of the model considered improved over the mean score of the reverberant utterances at $T_{60}=0.2s$; this was expected since a reverberation time of 0.2s represents a very subtle reverberation level. Reverberation times in [0.5, 1.0] seconds allow us to observe the dereverberation effectiveness of LS U-net and U-net GAN, since SRMR score is appreciably higher compared to reverberant utterances and the rest of models. U-net GAN (black) and the proposed architecture LS U-net (orange) show robust behavior in terms of reverberation time and also in terms of noise as previously shown in Table \ref{simulated1}.

\subsection{Qualitative evaluation for synthetic  data}

Figure \ref{qual_eval} shows an example of dereverberation performance. Note the similarity between clean spectrogram in Figure \ref{orig_example} and the three U-net variants output in \ref{unet_example}, \ref{gan_example} and \ref{LSunet_example}. LS U-net and U-net GAN architectures look visually identical.\\

\subsection{Results for real-world data}

\begin{table}[t]
\footnotesize
      \caption{SRMR results of LibriSpeech re-transmitted data for near and far microphones. Higher is better.}
       \centering  %% use this instead of \begin{center}
        \begin{tabular}{l c c c c c}
          \toprule
          {} & {\text{Near}} & {\text{Far}} \\\toprule[1pt]
          
          Reverberant & 3.99 & 4.36 \\
          Context-MLP & 4.69 & 5.53 \\
          Context-LSTM & 4.69 & 5.50 \\
          LS-LSTM & 5.49 & \textbf{8.16} \\
          FD-NDLP & 4.95 & 5.43 \\
          U-Net & 4.88 & 5.96 \\
          LS U-Net & 5.34 & 6.56 \\
          U-net GAN & \textbf{6.23} & 7.55 \\
          \bottomrule
        \end{tabular}
        \label{real_results}
        %\vspace{-2em}
\end{table}

Table \ref{real_results} shows the SRMR for the real-world data. U-net GAN exhibited the best results for near microphones and Late Reverberation Suppression LSTM (LS-LSTM) \cite{late_dev} for far microphones. Though the proposed architecture LS U-net did not exhibit the best performance for real data, it  improved over the baseline U-net and FD-NDLP performance for near and far microphones nonetheless. Critically, if we ranked the seven methods considered in Table \ref{real_results} based on their SRMR score, the proposed LS U-net would be third for both near and far microphones. This makes the proposed alternative for late reverberation suppression applied to U-net effective in real data.

%%%%%%%%%%%%%%%%%%%%%%%%%%%%%%%%%%%%%%%%%%%%%%%%%%%%%%%%%%%%%%%%%%%%

\section{Conclusions}
\label{conclusion}

We have proposed a U-net architecture for late reverberation suppression, termed LS U-net, and have experimentally validated it on synthetic and real-world data of different noise levels and reverberation times and microphone distances. Our results show that LS-Unet outperforms a wide range of deep-learning dereverberation methods under multiple performance indicators. In particular. LS U-net improves over the original U-net architecture and stands as a competitive alternative to the state-of-the-art GAN-trained extension of U-net. In the light of this results, future work wull focus on developing a GAN-trained version of the proposed LS U-net method.

\vspace{2em}

\noindent\textbf{Acknowledgements.} This work was funded by Fondecyt-Regular 1210606, ANID-AFB170001 (CMM) and ANID-FB0008 (AC3E).

\vfill\pagebreak
\bibliographystyle{IEEEbib}
\bibliography{refs}

\end{document}